\documentstyle[preprint,aps,epsf,floats,eqsecnum]{revtex}

\tighten

\begin{document}

\preprint{\vbox{\hbox{JHU--TIPAC--96006}
\hbox{TTP96--03}
\hbox{UTPT--96-02}
\hbox{hep-ph/9604290}}}

\title{Resumming Phase Space Logarithms \\
in Inclusive Semileptonic $B$ Decays}

\author{Christian Bauer$^{ab}$, Adam F. Falk$^c$ and Michael Luke$^b$}

\address{\medskip (a) Institut f\"ur Theoretische Teilchenphysik, Universit\"at
Karlsruhe\\ D-76128 Karlsruhe, Germany\\ \medskip
(b) Department of Physics, University of Toronto\\60 St.~George
Street, Toronto, Ontario,
Canada M5S 1A7\\ \medskip
(c) Department of Physics and Astronomy, The Johns Hopkins
University\\ 3400 North
Charles Street, Baltimore, Maryland 21218 U.S.A.\medskip}

\bigskip
\date{April 1996}

\maketitle

\begin{abstract}

We study logarithms of the form $\ln(m_q/m_b)$ which arise in
the inclusive semileptonic decay of a bottom quark to a quark of
mass $m_q$.  We use the renormalization group to resum the
leading radiative corrections to these terms, of the form
$m_q^2\alpha_s^n\ln^n(m_q/m_b)$,
$m_q^3\alpha_s^{n+1}\ln^n(m_q/m_b)$ and
$m_q^4\alpha_s^n\ln^{n+1}(m_q/m_b)$.  The first two resummations are trivial,
while the latter involves a non-trivial mixing of four-fermi operators in the
$1/m_b$
expansion.  We illustrate this technique in a toy model in which the
semileptonic decay
is mediated by a vector interaction, before treating the more complicated case
of
left-handed decay.
\end{abstract}

\pacs{13.20.He, 12.38.Bx, 13.20.Fc, 13.30.Ce}

\section{Introduction}

The inclusive semileptonic decay rate of a hadron containing a single bottom
quark may be written as a power series in
$\Lambda_{\rm QCD}/m_b$.  The leading term in this expansion is simply the
width for
the underlying quark-level process $b\rightarrow q\ell\bar\nu+q
g\ell\bar\nu+\dots$, which
may be expanded in powers of $\alpha_s(m_b)$,
\begin{equation}\label{rateexpand}
   \Gamma={G^2_F |V_{qb}|^2 m_b^5\over 192\pi^3}\left[\Gamma^{(0)}(\hat m_q)+
   {\alpha_s(m_b)\over \pi}\Gamma^{(1)}(\hat m_q)
   +\left({\alpha_s(m_b)\over \pi}
   \right)^2\Gamma^{(2)}(\hat m_q)+\dots\right].
\end{equation}
The coefficients $\Gamma^{(n)}$ are functions of the scaled final quark mass,
$\hat m_q = m_q/m_b$.  When the process is computed at the parton level, the
masses arise in
limits of the phase space integration, and hence the masses which appear are
the perturbatively defined pole masses, $m_q=m_q^{\rm pole}$.
Taking the leptons to be massless,  the tree level term is
\begin{equation}\label{treelevel}
   \Gamma^{(0)}(\hat m_q)=1-8\hat m_q^2-24\hat m_q^4\ln\hat m_q
   + 8\hat m_q^6-\hat m_q^8\,.
\end{equation}
The full expression for $\Gamma^{(1)}(\hat m_q)$ has been
computed analytically\cite{Nir}, and is quite lengthy; expanding the result in
powers of
$\hat m_q$ gives
\begin{eqnarray}\label{oneloop}
   \Gamma^{(1)}(\hat m_q)&=&
   {25\over6} - {2\pi^2\over 3}-\left({136\over3}+32\ln\hat m_q
   \right)\hat m_q^2+{64\pi^2\over 3}\hat m_q^3\nonumber\\
   &&-\left(182+{32\pi^2\over 3}-48\ln\hat m_q
   +96\ln^2\hat m_q\right)\hat m_q^4\\
   &&\mbox{}+{64\pi^2\over 3}\hat m_q^5
   -\left({2104\over27}-{608\over9}\ln\hat m_q\right)\hat m_q^2\\
   &&\mbox{}-\left({8857\over2700}+{2\pi^2\over 3}-{32\over15}\ln\hat m_q
   +{8\over3}\ln^2\hat m_q\right)\hat m_q^8
   +{\cal O}(\hat m_q^{10})\,.
\end{eqnarray}
At order $\alpha_s^2$, only the graphs containing gluon vacuum polarization
have
been calculated, and only numerically.  For example \cite{LSWpqcd}
\begin{eqnarray}\label{twoloop}
   \Gamma^{(2)}(0)&=&(-3.44\beta_0+c_1)\,,\nonumber\\
   \Gamma^{(2)}(0.37)&=&(-0.75\beta_0+c_2)\,,
\end{eqnarray}
where $c_1$ and $c_2$ denote terms not proportional to $\beta_0=11-2n_f/3$.

Since inclusive
semileptonic bottom decays provide a means of measuring the CKM mixing angle
$|V_{cb}|$, it is useful to have as much information as possible about the size
of the higher
order corrections to $\Gamma$.   In this paper, we use the operator product
expansion and the
renormalization group to sum to all orders leading logarithms of the form
$\hat m_q^4\alpha_s^n\ln^{n+1}\hat m_q$, for $n\ge0$, as well as terms of the
form
$\hat
m_q^2\alpha_s^n\ln^n\hat m_q$ and $\hat m_q^3\alpha_s^{n+1}\ln^n\hat m_q$.  We
will
show that these corrections are straightforward to calculate using the
renormalization
group.

We will see that the resummation of the ``phase space'' logarithms $\hat
m_q^4\alpha_s^n\ln^{n+1}\hat m_q$ is particularly interesting, and it is to
them
that we will pay
the most attention in what follows.  However, because of the prefactor $\hat
m_q^4$,
these terms are not
dominant as $\hat m_q\to0$, or in any other limit of the theory.  In fact, they
are smaller, in
principle, than uncomputed terms of the form $\hat
m_q^2\alpha_s^{n-1}\log^n\hat
m_q$, since
$\hat m_q\log\hat m_q\to0$ as $\hat m_q\to0$.  On the other hand, since $\hat
m_c$
is not particularly small, these terms may be numerically significant for
$b\rightarrow c$ decays.

Unfortunately, by the same token $\hat m_c\sim 0.37$ is such a poor expansion
parameter\footnote{For our numerical results, we use the HQET relation
$\hat m_c=\overline{m}_D/\overline{m}_B+{\cal O}(1/m_{b,c})$, where
$\overline{m}_M=
(m_M+3m_M^*)/4$ is the spin-averaged meson mass.  This gives $\hat m_c=0.37$
instead of
the more commonly used value of 0.3.}
that the terms which we can compute using the renormalization group do not
dominate those which we cannot compute as easily, and so these results may not
be
used directly to estimate the size of the higher order corrections to
$b\rightarrow X_c\ell\bar \nu$ decays.   This is
clear from examining the sizes of the various terms which contribute to
$\Gamma^{(0)}$ and $\Gamma^{(1)}$:
\begin{eqnarray}
   \Gamma^{(0)}(0.37) &=& 1-1.095+0.448+0.021-0.0004 +\dots = 0.372 \\
   \Gamma^{(1)}(0.37) &=& -2.41-1.85+10.67-8.06+1.46-0.372-0.005+\dots
=-0.568.\nonumber
\end{eqnarray}
where the order of the terms is the same as in Eqs.~(\ref{treelevel}) and
(\ref{oneloop}), and we have included terms up to ${\cal O}(\hat m_c^8)$ in the
expression for $\Gamma^{(1)}$.  Significant cancelations occur in both
expressions between
terms of different order in
$\hat m_c$; in particular, there is a large cancelation between the
${\cal O}(\alpha_s\hat m_c^3)$ and ${\cal O}(\alpha_s\hat m_c^4)$ terms.
Similarly, we will find when expanding our resummed results that there is a
large
contribution (larger than the tree level rate!) at
${\cal O}\left(\alpha_s^2\hat m_c^3\ln\hat m_c\right)$.  In analogy with the
one-loop
expression, we might expect a large cancelation between this term and the order
$\alpha_s^2\hat m_c^4\ln\hat m_c$ term.  However, this latter term is down by
two
powers of $\ln\hat m_c$ relative to the terms which we are resumming (requiring
a
three-loop anomalous dimension to resum), and so
we have not calculated it.   The motivation for our analysis is the insight it
will afford us into
the origin of a variety of higher order terms in the expression for the
semileptonic
width, rather than in any reliable estimate of the true size of higher order
corrections.

We will use the operator product expansion (OPE) and the heavy quark effective
theory (HQET)~\cite{VS,PW,Wisgur,EH1,Grinstein,Georgi}
in our analysis.  The application of OPE techniques to inclusive semileptonic
heavy quark decays
was suggested in Refs.~\cite{CGG,SV}, in which two distinct, but
ultimately equivalent,
methods were introduced.  The two approaches differ in the treatment of the
leptons in the final
state.  Since the leptons interact only weakly and electromagnetically with the
quark currents which
mediate the hadronic decay, there is freedom to integrate over their momenta at
various
stages of the calculation.

\begin{figure}
\epsfxsize=14cm
\hfil\epsfbox{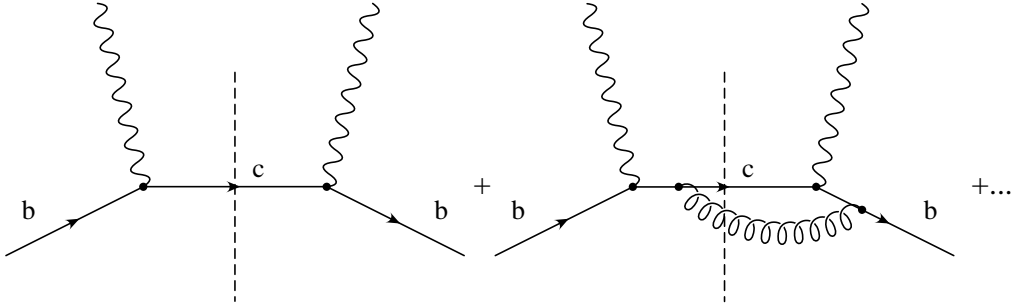}\hfill
\caption{Typical diagrams contributing to ${\rm Im}\,
T\{J^{\mu\dagger},J^\nu\}$.}
\label{ope1}
\end{figure}

Let us consider the decay $B(P_B)\to X_c(P_X)+(\ell\bar\nu)(q)$.  This process
is mediated by a term in the weak Hamiltonian,
\begin{equation}
  {\cal H}_W=\dots+{G_FV_{cb}\over\sqrt2}\,
  \overline c\gamma^\mu 1-\gamma_5)b\,
  \overline{\ell} \gamma_\mu (1-\gamma_5)\nu\,.
\end{equation}
In the
approach of
Ref.~\cite{CGG}, the Hamiltonian is explicitly factorized into a product of a
quark current, $J^\mu_h$ and a lepton current $J^\mu_\ell$.  Then the
differential rate ${\rm d}\Gamma/{\rm d} q^2\,{\rm d}
q\cdot v$ is
written as
\begin{equation}\label{contour}
   {{\rm d}\Gamma_{B\to X_c\ell\nu}\over {\rm d}q^2\,{\rm d}
   q\cdot v}\sim L_{\mu\nu}(q^2,q\cdot v)
   W^{\mu\nu}(q^2,q\cdot v)\,,
\end{equation}
where $v^\mu=P_B^\mu/m_B$ is the velocity of the $B$ meson and
$L_{\mu\nu}$ is the spin summed lepton tensor
($L_{\mu\nu} \propto ( q_\mu q_\nu - g_{\mu\nu} q^2 )$ for massless leptons.)
The nonperturbative hadronic tensor
$W^{\mu\nu}$ is  related via the optical theorem to the imaginary part of the
forward
scattering amplitude~\cite{CGG,SV},
\begin{eqnarray}
   W^{\mu\nu} &=& \sum_{X_c} \langle B|\,J_h^{\mu\dagger}\,|X_c\rangle
   \langle X_c|\,J_h^\nu\,|B\rangle (2\pi)^4\delta^4(P_B - P_X-q)\nonumber\\
   &=& -2\,{\rm Im}\,\langle B|\,i\int{\rm d}x\,e^{-iq\cdot x}
   \,T \left\{  J_h^{\mu\dagger}(x), J_h^\nu(0)  \right\}\,|B\rangle\,.
\end{eqnarray}
The time-ordered product is then written via an  operator product
expansion as a power series in $\alpha_s(m_b)$ and $1/m_b$, as illustrated in
Fig.~\ref{ope1}.  The resulting expression for the differential rate consists
of a series of delta
functions and derivatives of delta functions at the threshold for $c$ quark
production,
followed by a cut in the complex $q\cdot v$ plane corresponding to gluon
bremsstrahlung.
When integrated over the appropriate variables, this yields a sensible
prediction for
differential decay widths as well as for the total semileptonic width.
In this approach, the factor of $\hat m_c^4\ln\hat m_c$ in
Eq.~(\ref{treelevel})
arises from the
integration over the phase space variables $ q$ and $q\cdot v$.  It is
important
to
note that this term is not related to the running of the operators in the OPE
between $\mu=m_b$ and $\mu=m_c$, since this running is performed before
any phase space integrals are performed.

\begin{figure}
\epsfxsize=14cm
\hfil\epsfbox{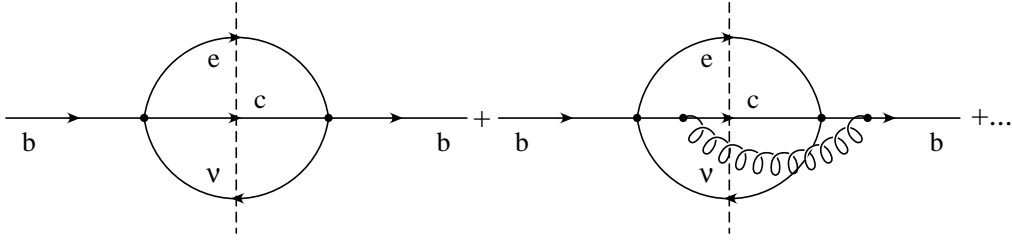}\hfill
\caption{Diagrams contributing to ${\rm Im}\,T\{{\cal H}_W^\dagger,
{\cal H}_W\}$.}
\label{ope2}
\end{figure}

By contrast, in the approach of Ref.~\cite{SV} the OPE is performed on the
expression
for the total, rather than the differential, rate.   Once again, the width is
written as the imaginary part of
the forward scattering amplitude,
\begin{eqnarray}\label{tproduct}
   \Gamma_{B\rightarrow X_c\ell\nu}&\sim& \sum_{X_c}{\rm Im}\,
  {\cal A}(B\rightarrow
   X_c e\bar\nu\rightarrow B)\nonumber \\
   &\sim&{\rm Im}\langle B\vert T\{{\cal H}_W^\dagger,{\cal H}_W\}
   \vert B\rangle\,.
\end{eqnarray}
This version of the time ordered product is  illustrated in Fig.~\ref{ope2}.
While this approach is completely equivalent to the other, performing the OPE
{\it after\/}
the integration over the lepton momenta gives us more insight into the origin
of the terms involving $\ln\hat m_c$.

\begin{figure}
\epsfxsize=6cm
\hfil\epsfbox{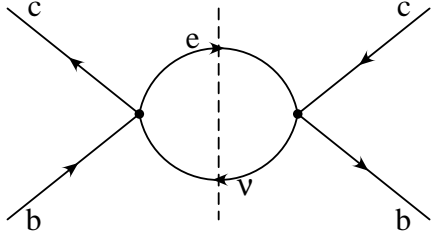}\hfill
\caption{Contribution to the coefficient function of $\overline b\Gamma_1
c\,\bar c\Gamma_2b$ in the OPE of ${\rm Im}\,T\{{\cal H}_W^\dagger,
{\cal H}_W\}$.}
\label{fullgraph}
\end{figure}

When the OPE is performed at the renormalization scale
$\mu=m_b$, the time ordered product in Eq.~(\ref{tproduct}) is written
as a sum of local operators,
\begin{eqnarray}
   \left.{\rm Im}\, T\{{\cal H}_W^\dagger,{\cal H}_W\}
   \right\vert_{\mu=m_b}&\longrightarrow &  a_1(m_b)\,\overline b\,
   b+{1\over m_b^2}\left\{ a_{21}(m_b) \hat m_c^2\,\overline b\, b+a_{22}(m_b)
   \overline b\,({\rm i D})^2\, b\right.\nonumber \\
   &&+\left. a_{23}(m_b) \overline b\,
   \sigma_{\mu\nu}G^{\mu\nu}\, b\right\}+{\cal O}\left({1/m_b^3}\right)
\end{eqnarray}
(note that only the imaginary part of the time ordered product is needed).
At the matching scale there are no factors of $\ln\hat m_c$ in the coefficient
functions $a_{ij}(m_b)$; these logarithms are infrared effects which are
contained
in the matrix elements of local operators.
However, since none of the operators at order $1/m_b^2$ contains explicit
$c$ quarks, none of their matrix elements depend on $m_c$ at leading order
in $\alpha_s$; therefore there is no term
proportional to $\hat m_c^2\ln\hat m_c$ in the tree level expression
(\ref{treelevel}) for the
semileptonic decay rate.  The first operators of interest which contain
explicit $c$
quarks are four quark operators of the form
$\overline b\Gamma_1 c\,\overline c\Gamma_2 b$.  These arise in the matching
due to
the graph in Fig.~\ref{fullgraph} and are of relative order $1/m_b^3$.

\begin{figure}
\epsfxsize=8cm
\hfil\epsfbox{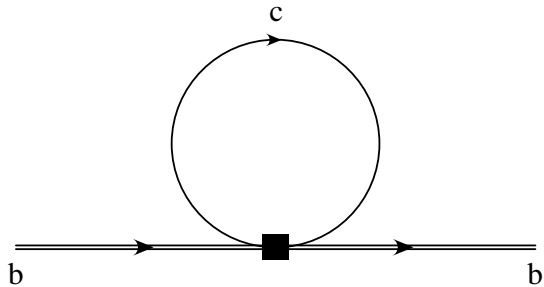}\hfill
\caption{Contracting the charm quark fields in a four fermion operator gives a
$\ln(m_c/m_b)$ contribution to the semileptonic width.}
\label{cloop}
\end{figure}

In the next section we will consider a toy model in which the $b$ quark decays
via a
vector current.  In this model, the operator $\bar h h\,\bar c c$ arises at
order $1/m_b^3$ in the OPE; its matrix element between $b$ quarks is given by
the diagram in Fig.~\ref{cloop}, giving a contribution of order
$\hat m_c^3\ln\hat m_c$ to the total inclusive rate.  For the physically
relevant
case of a left handed current the corresponding operator is $\bar h h\,
\bar c\rlap/v c$, for which the graph in
Fig.~\ref{cloop} vanishes.  This explains the lack of a term of order $\hat
m_c^3
\ln\hat m_c$ in the total inclusive rate~(\ref{treelevel}).   In this case, the
first logarithm of $\hat m_c$ arises at order $1/m_b^4$, in the matrix element
of the operator $\hat m_c\bar h h\,\bar c c$, giving the term of order
$\hat m_c^4\ln\hat m_c$ in the inclusive decay rate.

Rather than leave these logarithms in the matrix elements of local operators,
it is
convenient to scale the theory down to the renormalization point $\mu=m_c$,
at which point the $c$ quark is integrated out of the theory.  Below this scale
matrix
elements can no longer depend on $m_c$; all such dependence has been
transferred to
the coefficient functions in the OPE.  By including the leading QCD corrections
in the
renormalization group equation, the complete series of leading logarithms of
the form
$\hat m_c^n\alpha_s^n\ln^{n+1}\hat m_c$ may be resummed,
\begin{equation}\label{leadinglogs}
   c_0\hat m_c^n\ln\hat m_c\rightarrow\hat m_c^n\left(c_0
   \ln\hat m_c+c_1\alpha_s\ln^2\hat m_c+c_2\alpha_s^2\ln^3\hat
m_c+\dots\right)\,,
\end{equation}
where $n=3$ for a vector current and $n=4$ for a left handed current.
This calculation is technically more complicated for a left-handed current than
a
vector current, since it involves the renormalization of the complete set of
dimension
seven operators.  Therefore we will warm up in the next section with the
simpler case
of a vector current, before proceeding on to the realistic decay.

\section{Decays via a vector current}

We consider the decay mediated by the hadronic current $\overline c\gamma^\mu
b$, coupled to the usual left-handed massless leptons.  At tree level,
the decay width of the $b$ quark is given by
\begin{eqnarray}\label{vectortree}
   \Gamma_V&=&\Gamma_{V0}\big[1-2\hat m_c-8\hat m_c^2-18\hat m_c^3+18\hat m_c^5
   +8\hat m_c^6+2\hat m_c^7-\hat m_c^8\nonumber\\
   &&\mbox{}\qquad\qquad-24\hat m_c^3\log\hat m_c-24\hat m_c^4\log\hat m_c
   -24\hat m_c^5\log\hat m_c\big],
\end{eqnarray}
where $\Gamma_{V0}=m_b^5/192\pi^3$.
The total decay rate may be written via the optical theorem in terms of the
imaginary part of the forward scattering amplitude.  Integrating explicitly
over the leptons, this may be recast as the expectation value of the
time-ordered product of the hadronic current and its conjugate,
\begin{equation}\label{top}
   \Gamma_V = {1\over2m_B}\int {\rm d}q\,e^{-iq\cdot x}\,\langle B|\,
   T\{\overline b\gamma^\mu c(x),
   \overline c\gamma^\nu b(0)\}\,|B\rangle \times{1\over3\pi}
   \left(q_\mu q_\nu-q^2 g_{\mu\nu}\right)\,.
\end{equation}
The integral is taken over physical values of the total lepton four-momentum
$q^\mu$.

We now develop the time-ordered product in an operator product expansion.  The
momentum transfer is of order $m_b^2$ over almost the entire region
of integration, so we organize the expansion in inverse powers of $m_b$ rather
than in inverse powers of $q^2$.  Simultaneously, we will expand the ordinary
$b$ quark field in terms of the mass-independent HQET field $h$, defined
by~\cite{Georgi}
\begin{equation}\label{hdef}
   h(x) = {1+\rlap/v\over2}\,\exp(im_bv\cdot x)\,b(x)\,.
\end{equation}
Here $v^\mu=p_b^\mu/m_b$ is the four-velocity of the $b$ quark, which is fixed
in the limit $m_b\to\infty$~\cite{Wisgur}. The Dirac matrix $(1+\rlap/v)/2$
projects onto the
heavy quark part of the field operator, so $h$ is a two-component, rather than
a four-component, object.  The exponential factor cancels out the large
``on-shell'' part of the $b$ quark momentum.  This procedure will make all
dependence on the heavy quark mass $m_b$ explicit in the operator product
expansion.

We will order operators in inverse powers of $m_b$.  The operators of dimension
less than six can have no more than two fermion fields, which must be $h$ and
$\overline h$, since we are interested in the heavy quark decay process.  The
$c$ quark fields are contracted; the integral over the $c$ quark momentum
$p^\mu$ is equivalent to the integral over the lepton momentum
$q^\mu=p_b^\mu-p^\mu$ in Eq.~(\ref{top}).  The leading operator, then, is of
dimension three:
\begin{equation}\label{leadingvec}
   {m_b^5\over192\pi^3}\,\overline h\,h = \Gamma_{V0}\,\overline h\,h\,.
\end{equation}
The matrix element of $\overline h\,h$ may be expanded in inverse powers of
$m_b$~\cite{Wisgur},
\begin{equation}\label{matelvec}
   \langle B|\,\overline h\,h\,|B\rangle =
   2m_B\big[1+{\cal O}(1/m_b^2)\big]\,,
\end{equation}
where all $1/m_b^n$ corrections are independent of $\hat m_c$.  Hence the
leading
operator reproduces the $\hat m_c=0$ decay rate $\Gamma_{V0}$.

The next operator in the expansion in $1/m_b$ is of dimension four,
\begin{equation}\label{mcsquared}
   -2\Gamma_{V0}\,\hat m_c\,\overline h\,h\,.
\end{equation}
The factor of $\hat m_c$ is an explicit part of the operator, including its
dependence
on $\mu$.  The operator $\overline h\,h$ is a conserved current in the HQET and
does not run~\cite{FGGW}.  Since the operator product expansion is performed at
the scale
$\mu=m_b$, it is $m_c(m_b)$ which appears in the matching.  Since $m_b$ does
not run below $\mu=m_b$, neither does the combination
$\hat m_c(m_b)\overline h\,h$.
However,
leading logarithms are generated if the decay rate is expressed, as it usually
is,
in terms of $m_c(m_c)$ or $m_c^{\rm pole}$.  All of these
logarithms are resummed if we leave $m_c(m_b)$ unexpanded.  The situation is
analogous
for the dimension five operator
\begin{equation}\label{mccubed}
   -8\Gamma_{V0}\,\hat m_c^2\,\overline h\,h\,.
\end{equation}
Matching at $\mu=m_b$ yields $m_c(m_b)$ in Eqs.~(\ref{mcsquared}) and
(\ref{mccubed}), which resums
the leading logarithms of the form $\alpha_s^n\log^n\hat m_c$ in these terms.
We
note that there are other operators which arise at dimension five and higher,
such as $\overline h(iD)^2 h$ and $\overline h\sigma_{\mu\nu} G^{\mu\nu}h$,
but
the matrix elements of these operators are proportional to QCD scale quantities
such as $\lambda_1$ and $\lambda_2$ and do not yield powers of $\hat
m_c^n$~\cite{FN}.  They are included in an analysis which treats the
nonperturbative power corrections to
the parton model decay~\cite{Bigietc,MW,FLS,Mannel,FLNN}.

At dimension six we first encounter the four-fermion operators of the
form $\overline h\Gamma_1 c\,\overline c\Gamma_2 h$, which give rise at tree
level
to the term proportional to $\hat m_c^3\log\hat m_c$ in the total decay rate.
We
will now compute the leading logarithmic improvement of these terms.
Keeping the
leading powers of $m_b$ in the lepton tensor $(q_\mu
q_\nu-q^2g_{\mu\nu})/3\pi$, the
operator product expansion yields the dimension six operator
\begin{equation}
   \overline h\gamma^\mu c\,\overline c\gamma^\nu h\times{m_b^2\over3\pi}
   (v_\mu v_\nu-g_{\mu\nu})\,.
\end{equation}
With the identity $\rlap/v h=h$ on the HQET field, this reduces to
\begin{equation}\label{vecmatch}
   {m_b^2\over3\pi}\left[\overline hc\,\overline ch-
   \overline h\gamma^\mu c\,\overline c\gamma_\mu h\right].
\end{equation}
These operators are matched onto at the scale $\mu=m_b$.  Closing the charm
quark loop, they mix with the operator $\hat m_c^3\,\overline h\,h$ at order
$\alpha_s^0$, yielding the ``phase space'' logarithm $-24\Gamma_{V0}\hat m_c^3
\log\hat m_c$.  To resum the leading logarithms, we must consider the QCD
renormalization of the dimension six operators.  However, we do not
need the QCD correction to the mixing of the four quark operators with the
quark bilinears, which is subleading and will only produce terms in the rate
proportional
to $\hat m_c^3\alpha_s^n\log^n\hat m_c$.

The renormalization of the four quark operators simplifies considerably if we
first apply a Fierz transformation to bring them into the form $\overline
h\Gamma_1 h\,\overline c\Gamma_2 c$.  We use the $SU(3)$ identity
\begin{equation}
   \delta_{il}\,\delta_{kj} = 2\,T^a_{ij}\,T^a_{kl}+\case13\,
   \delta_{ij}\,\delta_{kl}\,,
\end{equation}
on the color indices, and include a factor of $-1$ from the exchange of the
fermion fields $h$ and $c$.  The Fierz transformation yields a linear
combination of operators of the form
\begin{equation}
   {\cal O}^1_a = \overline h\Gamma_a h\,\overline c\Gamma_a c\qquad
   {\rm and}\qquad
   {\cal O}^8_a = \overline h\Gamma_a T^ah\,\overline c\Gamma_a T^ac\,,
\end{equation}
where $\Gamma_a=1,\gamma^5,\gamma^\mu,\gamma^\mu\gamma^5,\sigma^{\mu\nu}$.
However, because the heavy field $h$ has only two components, there are
actually just four independent scalars of the form $\overline h\Gamma h$.  It
is straightforward to show, then, that the allowed Dirac structures for the
dimension six operators are
\begin{eqnarray}\label{dirac}
  &&\overline h\,h\,\overline c\,c\,,\nonumber\\
  &&\overline h\,h\,\overline c\rlap/v c\,,\nonumber\\
  &&\overline h\gamma^\mu\gamma^5 h\,\overline c\gamma^\mu\gamma^5c\,,\\
  &&\overline h\gamma^\mu\gamma^5 h\,
    \overline c\rlap/v \gamma^\mu\gamma^5c\,.\nonumber
\end{eqnarray}
With both singlet and octet color structures, we find a set of eight operators.

In the HQET, the coupling of a gluon to a heavy quark line is given by the
Feynman rule $v\cdot A$, where $A^\mu$ is the gluon field~\cite{Georgi}.  In
$v\cdot A=0$
gauge, in which the gluon does not couple directly to the heavy quark, we see
that the
renormalization of ${\cal O}^1_a$ and ${\cal O}^8_a$ is restricted to the
$\overline c\Gamma c$ part of the
operator.  Since for $\hat m_c=0$ the addition of a gluon loop to $\overline
c\Gamma c$ generates an {\it even\/} number of Dirac matrices and no
$\gamma^5$, none of the eight dimension six operators mix with each other under
renormalization.  Hence the running of these eight operators is multiplicative,
except for the color octet $\overline hT^ah\,\overline c
\rlap/v T^ac$, which mixes via ``penguin'' diagrams with the flavor $SU(3)$
singlet operator $\overline hT^ah\,\overline q_i\rlap/v T^aq_i$, with $i$
summed over $u,d,s$.

The dimension six operators mix with quark bilinears at order $\alpha_s^0$ by
contracting the charm quark fields as in Fig.~\ref{cloop}.  However, only the
color singlet,
scalar-scalar operator
\begin{equation}
   {\cal O}_6=\overline h\,h\,\overline c\,c
\end{equation}
has a color and Dirac structure such that this mixing, to $\hat m_c^3\overline
h\,h$, is nonvanishing.  Hence it is the only operator which we must consider.
The Fierz transformation of the operators (\ref{vecmatch}) yields the
coefficient function
\begin{equation}
   C_6(m_b) = {m_b^2\over12\pi}\,.
\end{equation}
$C_6(\mu)$ runs according to the renormalization group equation
\begin{equation}\label{rge1}
   \mu{{\rm d}\over{\rm d}\mu}\,C_6(\mu) = \gamma_6\,C_6(\mu)\,,
\end{equation}
where a simple HQET calculation yields (via the graphs in Fig.~\ref{fourfermi})
\begin{equation}
   \gamma_6 = -{g^2(\mu)\over2\pi^2}\,.
\end{equation}
Solving the differential equation (\ref{rge1}) by standard methods, we find
\begin{equation}
   C_6(\mu) = {m_b^2\over12\pi}\,
   \left[{\alpha_s(\mu)\over\alpha_s(m_b)}\right]^{12/25}\,.
\end{equation}

We now turn to the renormalization of the term $C_3(\mu)\hat
m_c^3(\mu)\overline h\,h$.
The coefficient $C_3(\mu)$ runs both because of the renormalization of
the operator
\begin{equation}
   {\cal O}_3 = \hat m_c^3\overline h\,h\,,
\end{equation}
and because of the mixing from ${\cal O}_6$.  It obeys the renormalization
group equation
\begin{equation}\label{rge2}
   \mu{{\rm d}\over{\rm d}\mu}\,C_3(\mu) = \gamma_3\,C_3(\mu)+
   \gamma_{63}\,C_6(\mu)\,.
\end{equation}
Since the current $\overline h\,h$ is conserved, the anomalous dimension
$\gamma_3$ is given solely by the renormalization of the mass $\hat
m_c(\mu)$,
\begin{equation}
   \gamma_3=3\gamma_m = -{3g^2(\mu)\over2\pi^2}\,.
\end{equation}
The mixing of ${\cal O}_6$ with ${\cal O}_3$ is given by contracting the charm
quark fields as in Fig.~\ref{cloop},
\begin{equation}
   \gamma_{63} = {3\over2\pi^2}\,m_b^3\,.
\end{equation}
Because the graph has a cubic divergence, it is proportional to $m_c^3$; hence
$\gamma_{63}$ is proportional to $m_b^3$.

Solving the differential equation (\ref{rge2}) for $C_3(\mu)$ and setting
$\mu=m_c$, we find
\begin{equation}\label{c3solve}
   C_3(m_c) = {m_b^5\over192\pi^3}\,\hat m_c^3{144\pi\over23\alpha_s(m_c)}\,
   \left[z^{12/25}-z^{-11/25}\right]\,,
\end{equation}
where
\begin{equation}
  z = {\alpha_s(m_c)\over\alpha_s(m_b)} > 1\,.
\end{equation}
Note that because the numerator in the expression (\ref{c3solve}) vanishes as
$z\to1$, the solution for $C_3(\mu)$ is well behaved in the limit
$\alpha_s(m_c)\to0$.  Expanding $C_3(\mu)$ in powers of $\alpha_s(m_c)$, we
find
\begin{equation}
   C_3(m_c) = \Gamma_{V0}\,\hat m_c^3\left[-24\ln\hat m_c
   -48{\alpha_s(m_c)\over\pi}\ln^2\hat m_c+\dots\right]\,.
\end{equation}
For $\alpha_s(m_c)=0.41$ and $\hat m_c=0.37$ (for which $\alpha_s(m_b)=0.27$
and
$z=1.54$), we find that the contribution of this operator to the total rate is
of order one:
\begin{equation}
   C_3(m_c) = \Gamma_{V0}\,[1.22-0.21+\dots]=\Gamma_{V0}\,[0.99]\,.
\end{equation}
The resummed logs change the total decay rate by
$18\%$.  In terms of the quantities $\hat m_c(m_b)$ and $\alpha_s(m_b)$
renormalized
at the scale $\mu=m_b$, we find
\begin{eqnarray}
   \Gamma=\Gamma_{V0}\bigg\{1&&-2\hat m_c(m_b)-8\hat m_c^2(m_b)
   -18\hat m_c^3(m_b)\ln\hat m_c\\
   &&\mbox{}+{144\pi\over23\alpha_s(m_b)}\hat m_c^3(m_b)
   \left(z^{23/25}-1\right)+\dots\bigg\}\,.
\end{eqnarray}
Note that although we include the term $-18\hat m_c^3(m_b)\ln\hat m_c$ in
this renormalization group improved expression, there are also uncomputed terms
of the
same order from the two-loop renormalization of the dimension six operators.

\section{Decays via a left-handed current}

We now apply the same analysis to the physical situation of decays mediated by
the
left-handed current $J^\mu=\overline b\gamma^\mu(1-\gamma^5)c$.  Including the
coupling $G_FV_{cb}/\sqrt2$, the total decay rate is related to the forward
scattering
amplitude via
\begin{equation}\label{topleft}
   \Gamma = {1\over2m_B}{G_F^2|V_{cb}|^2\over2}
   \int {\rm d}q\,e^{-iq\cdot x}\,\langle B|\,
   T\{J^{\mu\dagger}(x),J^\nu(0)\}\,|B\rangle
   \times{1\over3\pi}
   \left(q_\mu q_\nu-q^2 g_{\mu\nu}\right)\,.
\end{equation}
The tree level rate is $\Gamma_0=G_F^2|V_{cb}|^2m_b^5/192\pi^3$.  We now expand
the time ordered product in an operator product expansion, as before.  At tree
level,
and for dimension $n<6$, we find
 \begin{equation}
   {\Gamma_0\over2m_B}\left[\overline h\,h - 8\hat m_c^2(m_b)\overline h\,h\
   +\dots\right]\,,
\end{equation}
where the OPE is performed at the renormalization scale $\mu=m_b$.  The HQET
field $h(x)$ is given in Eq.~(\ref{hdef}), and the lowest order matrix element
of $\overline h\,h$ in Eq.~(\ref{matelvec}).   The ellipses denote
charm-independent dimension five
operators such as $\overline h({\rm iD})^2h$, which do not induce terms
proportional to
$\hat m_c^n$.  As before, the combination $\hat m_c^2\overline h\,h$ does not
run
below
$\mu=m_b$; expanding $\hat m_c(m_b)$ in terms of $m_c^{\rm pole}$, one finds at
leading logarithmic order
\begin{equation}
   -8m_c^2(m_b) = -8(m_c^{\rm pole})^2\,z^{-24/25} = (m_c^{\rm pole})^2
   \left[-8-32\alpha_s(m_b)\ln\hat m_c+{8\over 3}\alpha_s(m_b)^2\ln^2\hat
   m_c+\dots\right]\,,
\end{equation}
where $z=\alpha_s(m_c)/\alpha_s(m_b)$.  Hence we reproduce the order
$\alpha_s$ correction from $\Gamma^{(1)}$ (\ref{oneloop}), and then extend this
result to
resum all logarithms of the form $\hat m_c^2\alpha_s^n\ln^n\hat m_c$ in the
coefficient
functions $\Gamma^{(n)}$.  (This constraint on the $m_c^2\alpha_s\log\hat m_c$
was
also noted by Nir~\cite{Nir}).

At dimension six, four quark operators of the form
$\overline h\Gamma _1h\,\overline c\Gamma_2 c$ arise, just as in the case of
vector decays.  In principle, these operators
could induce terms of order $\hat m_c^3\ln\hat m_c$ in the total decay rate,
when
they mix with $\overline h\,h$.
Expanding the operator product, applying the Fierz transformation, and dropping
parity-odd operators which cannot contribute to the forward matrix element, we
find the dimension six operators
\begin{equation}
   {\Gamma_0\over2m_B}{32\pi^2\over m_b^3}\left[
   -{1\over2}\overline h\,h\,\overline c\rlap/vc
   -{1\over6}\overline h\gamma^\mu\gamma^5h\,\overline c\gamma^\mu\gamma^5c
   -3\overline hT_ah\,\overline cT_a\rlap/vc
   -\overline hT_a\gamma^\mu\gamma^5h\,\overline cT_a\gamma^\mu\gamma^5c
   \right].
\end{equation}
Note the absence of a term proportional to $\overline h\,h\,\overline c\,c$,
the only dimension six
operator which can mix with $\hat m_c^3\overline h\,h$.  Hence, unlike the case
of vector
decay, there is no term in the decay rate proportional to $\hat m_c^3\ln\hat
m_c$.
In fact,
since none of the dimension six operators of the form $\overline h\Gamma
h\,\overline
c\Gamma c$ and $\overline hT_a\Gamma h\,\overline cT_a\Gamma c$ mix with each
other (as discussed in the previous section), there are no terms in the decay
rate
proportional to $\hat m_c^3\alpha_s^n\ln^{n+1}\hat m_c$, for any $n$.  The
absence
of such
logarithmic terms at tree level in $\Gamma^{(0)}$ is thus extended to all
orders.   The leading logarithms at order $\hat m_c^3$ are therefore simply
resummed by replacing $m_c^{\rm pole}$ with $m_c(m_b)$ in the expression
for $\Gamma^{(1)}$,
\begin{equation}
  {64\pi^2\over 3}\,m_c(m_b)^3={64\pi^2\over 3}\,
  (m_c^{\rm pole})^3\,z^{-36/25}=
  {64\pi^2\over 3}\,(m_c^{\rm pole})^3\,
  \bigg\{1+6\alpha_s(m_b)\ln\hat m +\dots\bigg\}\,.
\end{equation}

To reproduce the $\hat m_c^4\ln\hat m_c$ term in $\Gamma^{(0)}$, we must
continue
the
OPE to include operators of dimension seven.  Although there is a large number
of such
operators, we can use heavy quark symmetry and the classical equations to
motion to reduce these to only a few which are relevant to the analysis.
There are three classes of dimension seven operators, each of which can have a
singlet
or octet color structure and one of the four Dirac structures (\ref{dirac}).
The first class
is dimension six operators multiplied by an additional factor of $\hat m_c$.
Because we
are counting powers of $1/m_b$, we will treat these operators as dimension
seven.
These operators do not mix with each other, just as their dimension six
counterparts do
not.  Hence the only one of these operators which can mix with $\hat
m_c^4\overline h\,h$ is $\hat m_c\overline h\,h\,\overline c\,c$.

\begin{figure}
\epsfxsize=14cm
\hfil\epsfbox{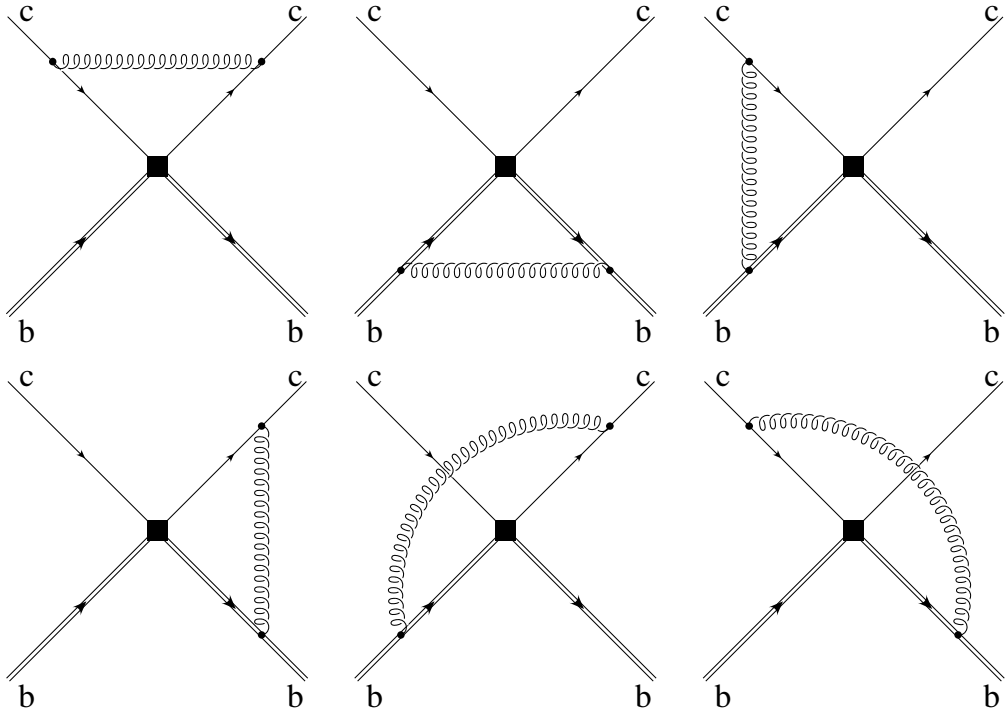}\hfill
\caption{Diagrams which renormalize ${\cal O}_1,\ldots,{\cal O}_6$.}
\label{fourfermi}
\end{figure}

\begin{figure}
\epsfxsize=14cm
\hfil\epsfbox{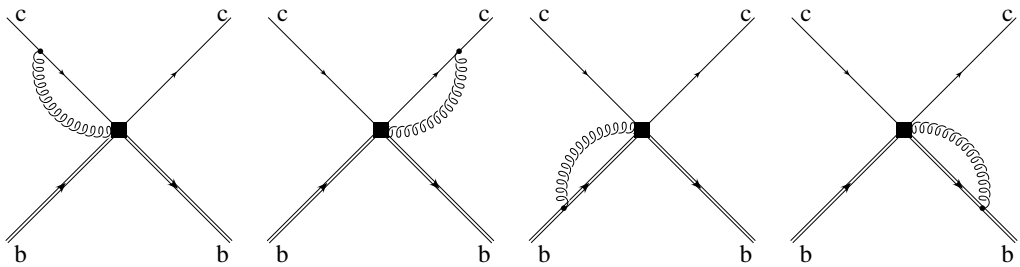}\hfill
\caption{Additional diagrams which renormalize ${\cal O}_3$, ${\cal O}_4$ and
${\cal O}_5$.}
\label{fourfermib}
\end{figure}

The second class of operators is those in which a derivative acts on the charm
quark.  We may use the classical equation of motion ${\rm i}\rlap{\,/}{\rm
D}c(x)=m_c\,c(x)$ to reduce some of these to operators of the first class.  Of
those that remain, only $\overline h\,h\,\overline c\rlap/v v\cdot {\rm iD}c$
mixes with $\hat
m^4\overline h\,h$, via the graphs in Fig.~\ref{cloop}.  However, these
operators can mix
with each other under renormalization, as well as with those of the first
class, via the
diagrams in Figs.~\ref{fourfermi} and \ref{fourfermib}.  Let us consider the
gauge in
which $v\cdot A=0$, where the gluon does not couple to the heavy quark field.
Then
we have only the graphs which renormalize the charm quark part of the operator.
 These can mix neither Dirac nor color structures, nor can they induce
operators in which a derivative acts on the bottom quark.  The running of these
operators is diagonal as well.  Finally, of the operators of this class, only
$\overline h\,h\overline c\rlap/v v\cdot {\rm iD}c$ mixes with $\hat
m\overline h\,h\,\overline c\,c$.

The third class is those operators in which a derivative acts on the bottom
quark.
Here the classical equation of motion $v\cdot{\rm iD}h(x)=0$ eliminates some
operators entirely.  The remaining ones
cannot mix directly with $\hat m^4\overline h\,h$ at order $\alpha_s^0$, but
they can
mix with the two other dimension seven operators we have identified so far.  In
$v\cdot
A=0$ gauge, such mixing can only occur via the first two graphs in
Fig.~\ref{fourfermib}.  Only color octet operators can mix with the color
singlet
$\overline h\,h\overline c\rlap/v v\cdot {\rm iD}c$ by such one-gluon exchange.
The operators in this class can also mix among themselves.  However, it is
straightforward to show that while color octets mix with both singlets and
octets, color
singlets mix only with each other.  Hence, in this class of operators, only the
color octets
are relevant to our calculation.  Furthermore, the Dirac structure is
sufficiently
constraining that of these, only $\overline h T_a {\rm iD}_\mu h\,\overline
cT_a\gamma^\mu c$ contributes.

\begin{figure}
\epsfxsize=14cm
\hfil\epsfbox{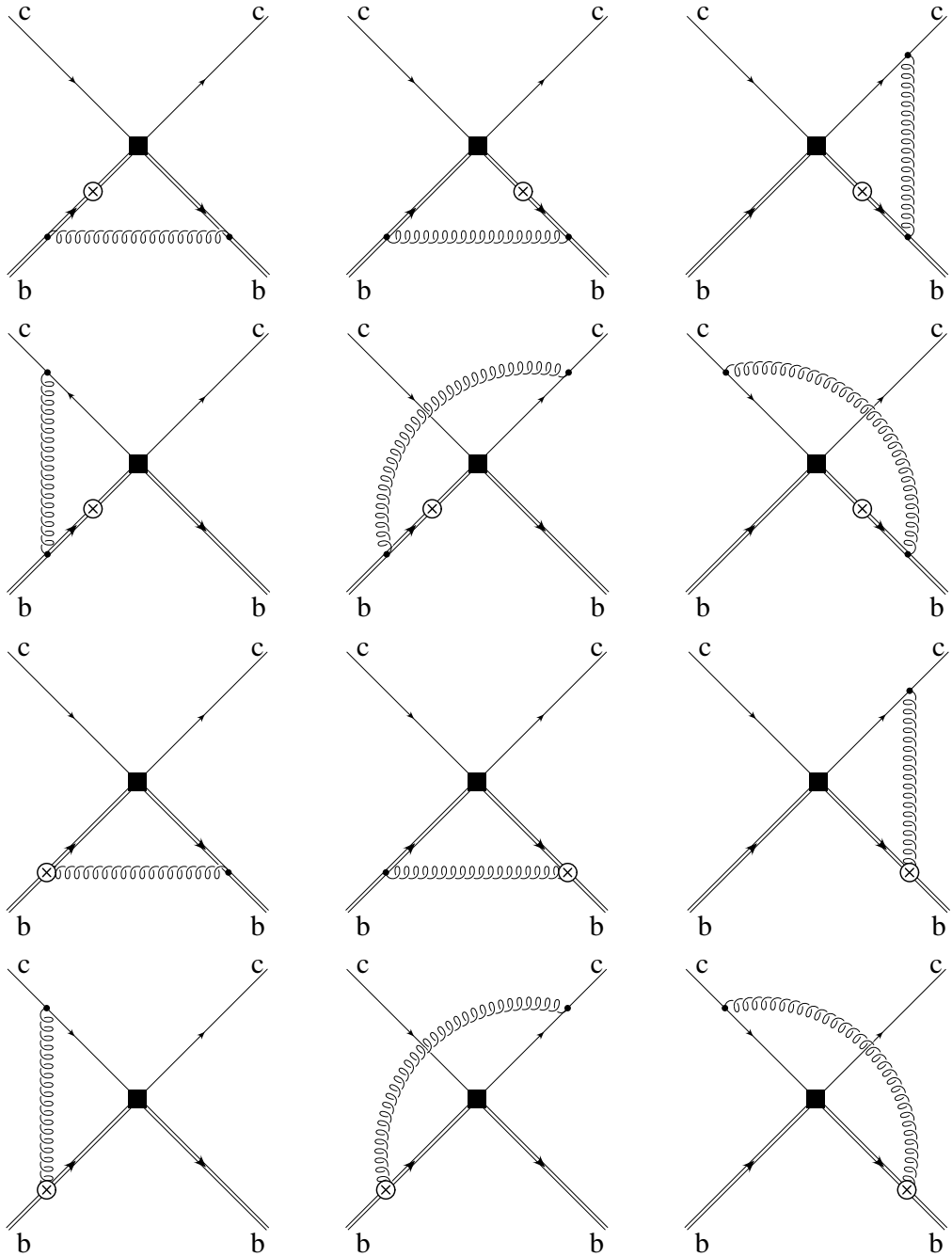}\hfill
\caption{Diagrams with a single insertion of ${\cal O}_K$
mixing ${\cal O}_1$ with operators of dimension 7.}
\label{insertions}
\end{figure}

There is one other source of dimension seven operators which we must consider.
The
HQET lagrangian beyond leading order is given by~\cite{FGL,EH2}
\begin{equation}
   {\cal L}_{\rm HQET} = \overline hv\cdot{\rm iD}h
   +{1\over m_b}{\cal O}_K+{1\over m_b}C_G(\mu){\cal O}_G(\mu)+\dots\,,
\end{equation}
where
\begin{equation}
   {\cal O}_K={1\over2}\,\overline h({\rm iD})^2 h
   \qquad{\rm and}\qquad
   {\cal O}_G={g\over4}\,\overline h\sigma^{\mu\nu}G_{\mu\nu} h
\end{equation}
are the leading spin and flavor symmetry violating corrections to the
$m_b\to\infty$
limit.  The operators ${\cal O}_K$ and ${\cal O}_G$ are treated as
perturbations; because
they come with explicit factors of $1/m_b$, they can induce mixing between
operators at different order in the $1/m_b$ expansion~\cite{FG}.  In
particular, they can mix operators of dimension six
with those of dimension seven, via the graphs shown in Fig.~\ref{insertions}.
We find
that insertions of ${\cal O}_G$ do not induce mixing with any of the dimension
seven
operators of interest, while ${\cal O}_K$ does, if the dimension six operator
which is being
renormalized is $\overline hT_ah\,\overline c\rlap/vT_ac$.

\begin{figure}
\epsfxsize=14cm
\hfil\epsfbox{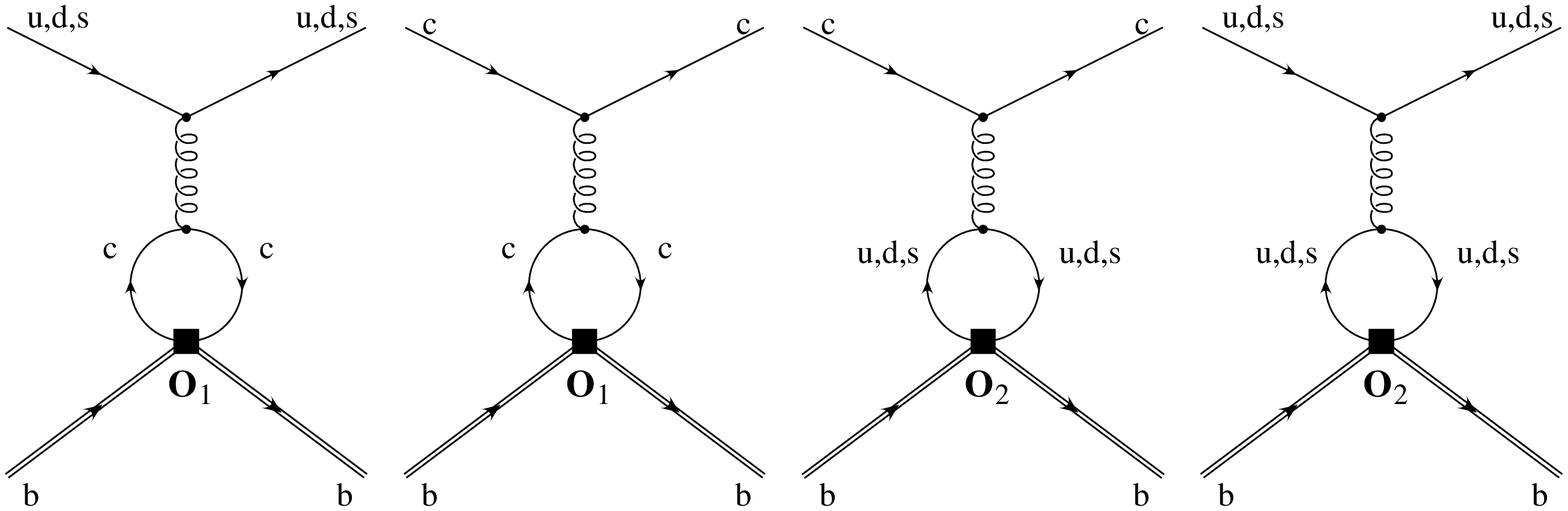}\hfill
\caption{``Penguin" diagrams mixing ${\cal O}_1$ and ${\cal O}_2$.  The same
diagrams mix ${\cal O}_3$ and ${\cal O}_4$.}
\label{penguin}
\end{figure}

With this taxonomy in hand, we can now write down the list of
operators of
dimension six and seven which are relevant for our analysis.  In the end, it is
mercifully short:
\begin{eqnarray}\label{oplist}
   {\cal O}_1&=& \overline h T_a h\;\overline c\rlap/v T_a c\nonumber \\
   {\cal O}_2&=& \overline h T_a h\;\overline q_i\rlap/v T_a q_i\,,
   \qquad q_i=u,\,d,\,s\nonumber \\
   {\cal O}_3&=&{1\over 2m_b} \overline h\big\{
   {\rm -i}\overleftarrow {\rm D}\!_\mu
   T_a+T_a{\rm i}\overrightarrow {\rm D}\!_\mu\big\} h\;
   \overline c \gamma^\mu T_ac\nonumber \\
   {\cal O}_4&=&{1\over 2m_b} \overline h\big\{
   {\rm -i}\overleftarrow {\rm D}\!_\mu
   T_a+T_a{\rm i}\overrightarrow {\rm D}\!_\mu\big\} h\;
   \overline q_i \gamma^\mu T_aq_i\,,\qquad q_i=u,\,d,\,s\\
   {\cal O}_5&=&{1\over 2m_b}\overline h\,h\;\overline c\rlap/v v^\mu
   \big\{{\rm -i\overleftarrow {\rm D}}\!_\mu
   +{\rm i\overrightarrow {\rm D}}\!_\mu\big\} c\nonumber \\
   {\cal O}_6&=&{m_c(\mu)\over m_b}\,\overline h\,h\;\overline c\,c\nonumber
\end{eqnarray}
The operators ${\cal O}_3$, ${\cal O}_4$ and ${\cal O}_5$ have been constructed
to be
Hermitian.  We
also must include the quark bilinear $m_c^4\overline h\,h$.  It is convenient
to define
it with an inverse factor of the strong coupling, because this will make the
anomalous
dimension matrix homogeneous in $g^2$:
\begin{equation}
   {\cal O}_7 = {1\over g^2(\mu)}\,{m_c^4(\mu)\over m_b}\,\overline h\,h\,.
\end{equation}
Factors of $1/m_b$ have also been included in  ${\cal O}_3,\ldots,{\cal O}_7$
so that the anomalous dimension matrix will have no explicit factors of
$1/m_b$.\footnote{We did not include these factors in the analysis of the
vector decay, because the renormalization group equations were already so
simple.}
To summarize, the operators ${\cal O}_1$ through ${\cal O}_6$ are renormalized
via the graphs in
Fig.~\ref{fourfermi}.   In addition, the dimension seven operators ${\cal
O}_3,\ldots,{\cal O}_6$ get contributions from the graphs in
Fig.~\ref{fourfermib}. The dimension
six
operators ${\cal O}_1$ and ${\cal O}_2$ mix with each other via the ``penguin''
diagrams in Fig.~\ref{penguin}, as do ${\cal O}_3$ and ${\cal O}_4$.  Finally,
${\cal O}_1$ mixes with ${\cal O}_5$ and ${\cal O}_6$ via time-ordered products
with
${\cal O}_K$, as shown in Fig.~\ref{insertions}.

The operators in Eq.~(\ref{oplist}) are in fact not all independent; ${\cal
O}_1$ and
${\cal O}_3$ (and ${\cal O}_2$ and ${\cal O}_4$) are related via
reparameterization invariance\cite{reparam}.  Since
$v_\mu$ and
${\rm D}_\mu$ must appear in the combination
\begin{equation}
  {\cal V}_\mu= v_\mu+i\overrightarrow{\rm D}_\mu/2m_b-i
  \overleftarrow{\rm D}_\mu/2m_b\,,
\end{equation}
we find the restrictions
\begin{equation}
  C_1(\mu)=C_3(\mu),\ \ C_2(\mu)=C_4(\mu)\,.
\end{equation}
Our explicit calculations confirm this result.

We now perform the operator product expansion at the scale $\mu=m_b$.  For the
operators of dimension seven, the matching coefficients are generated at
subleading
order in the expansion in $1/m_b$.   The momentum of the heavy
$b$ quark is written as $p_b^\mu=m_bv^\mu+k^\mu$, where $k^\mu$ is the
``residual'' momentum.  For an on-shell $h$ field, the classical equation of
motion is
$v\cdot k=0$~\cite{Georgi}.  The expression of the $b$ quark spinor $u_b$ in
terms of the heavy
spinor $u_h$ is also affected, becoming
$u_b=(1+\rlap/k/2m_b)u_h$~\cite{FGL,FG,Luke}.

There are then two sources of matching onto operators of dimension seven.
First,
the lepton momentum $q^\mu$ may be written as $q^\mu=p_b^\mu-p_c^\mu=
m_bv^\mu+k^\mu-p_c^\mu$.  The momenta $k^\mu$ and $p_c^\mu$ lead to
operators with covariant derivatives acting on the $h$ and $c$ fields,
respectively.
Second, the correction to the heavy quark spinors must be accounted for.  We
define
reduced operator coefficients by
\begin{equation}
   \hat C_i(\mu)={32\pi^2\over m_b^3}{\Gamma_0\over2m_B}\,C_i(\mu)\,.
\end{equation}
Then performing the operator product expansion at tree level yields the nonzero
terms
\begin{equation}\label{matching}
   \hat C_1(m_b)=-3\,,\qquad \hat C_3(m_b)=-3\,,\qquad
   \hat C_5(m_b)={2\over3}\,,\qquad\hat C_6(m_b)={1\over3}\,.
\end{equation}
The Wilson coefficients evolve according the renormalization group equation
\begin{equation}\label{rge}
   \mu{{\rm d}\over{\rm d}\mu}\hat C_i(\mu)=\gamma_{ji} \hat C_j(\mu)\,.
\end{equation}
The anomalous dimension matrix $\gamma_{ij}$ is defined by the operator
renormalization
\begin{equation}\label{defan}
   \gamma_{ij}\Gamma^{(n)}_{{\cal O}_j}=-\left(\mu{\partial\over\partial\mu}
   +\beta{\partial\over\partial\beta}+\gamma_{m_c} m_c{\partial\over
   \partial m_c}-n\gamma_{\rm ext}\right)\Gamma^{(n)}_{{\cal O}_i}\,,
\end{equation}
where $\Gamma^{(n)}_{{\cal O}_i}$ is an $n$-point Green function with a single
insertion of
the operator ${\cal O}_i$.   Using the known mass and wavefunction anomalous
dimensions~\cite{VS,PW,EH1,FGGW}
\begin{equation}
   \gamma_{m_c}=-{g^2\over2\pi^2}\,,\qquad
   \gamma_c={g^2\over 12\pi^2}\,,\qquad\gamma_h=-{g^2\over 6\pi^2}\,,
\end{equation}
the Feynman diagrams in Figs.~\ref{cloop}--\ref{penguin} yield the anomalous
dimension matrix
\begin{equation}\label{andim}
   \gamma_{ij}={g^2\over\pi^2}\left(
   \matrix{{-23/48}&{1/12}&0&0&-{2/9}&-{1/9}&0\cr
          {1/4}&-{5/16}&0&0&0&0&0\cr
          0&0&{-23/48}&{1/12}&0&{1/6}&0\cr
          0&0&{1/4}&-{5/16}&0&0&0&\cr
          0&0&0&0&{4/9}&-{1/9}&-{3/8}\cr
          0&0&0&0&0&0&-{3/2}\cr
          0&0&0&0&0&0&{23/24}\cr}\right).
\end{equation}
The renormalization group is then used to evolve the coefficients $\hat
C_i(\mu)$
from $\mu=m_b$ to $\mu=m_c$.  The logarithmically enhanced terms $\hat
m_c^4\alpha_s^n\ln^{n+1}\hat m_c$ are given by the combination
\begin{equation}
   F(m_c) = C_7(m_c)\,{m_c^4(m_c)\over g^2(m_c)m_b}
   \langle B|\,\overline h\,h\,|B\rangle
   = {8\pi\over\alpha_s(m_c)}\,\hat m_c^4\,\hat C_7(m_c)\,
   {\Gamma_0\over2m_B}\langle B|\,\overline h\,h\,|B\rangle\,.
\end{equation}
By inspection of the matching coefficients~(\ref{matching}) and the anomalous
dimension matrix~(\ref{andim}), we see that only
the linear combination ${\cal O}_5-4{\cal O}_6$ mixes into ${\cal O}_7$, and
that this linear combination
of operators does not run in the leading logarithmic approximation.  Therefore,
the solution to the renormalization group equation for $\hat C_7(\mu)$ is
particularly
simple.
With the matrix element (\ref{matelvec}), we find
\begin{equation}
   F(m_c) = \Gamma_0\,{8\pi\over\alpha_s(m_c)}\,\hat m_c^4\,
   {18\over23}\,(1-z^{-23/25})\,.
\end{equation}
When expanded and written in terms of $\alpha_s(m_b)$ and the reduced pole mass
$\hat m_c$, the
first two  terms of the expression
\begin{eqnarray}
   \Gamma_0^{-1}\, F(m_c) &=& -24\hat m_c^4\ln\hat m_c
   - 96 {\alpha_s(m_b)\over\pi}\,\hat m_c^4\ln^2\hat m_c +
   {32\over 3}\left({\alpha_s(m_b)\over\pi}\right)^2\hat m_c^4\ln^3\hat m_c
   \nonumber \\
   &&\mbox{}-12\left({\alpha_s(m_b)\over\pi}\right)^3\hat m_c^4\ln^4\hat m_c
   +{104\over5}
   \left({\alpha_s(m_b)\over\pi}\right)^4\hat m_c^4\ln^5\hat m_c+\dots\,,
\end{eqnarray}
reproduce the known tree level and one loop results from Eqs.~(\ref{treelevel})
and
(\ref{oneloop}), respectively.  In addition, all of the leading logarithms have
been
resummed.

As in the previous section we use $\alpha_s(m_c)=0.41$ and $\hat m_c=0.37$
to find
\begin{equation}
	F(m_c)=\Gamma_0\,[0.45-0.15-1.4\cdot10^{-3}+\ldots]=\Gamma_0\,[0.30]
\end{equation}
The re-expanded contributions from the terms $\hat m_c^2\overline{h}h$ and
$\hat m_c^3\overline{h}h$ are
\begin{equation}
  {\Gamma_0\over2m_B}\,\hat m_c^2(m_b)\langle B|\,\overline h\,h\,|B\rangle=
  \Gamma_0\,[-1.10-0.37-0.12+\ldots]=\Gamma_0\,[-1.67]
\end{equation}
and
\begin{equation}
  {\Gamma_0\over2m_B}\,\hat m_c^3(m_b)\langle B|\,\overline h\,h\,|B\rangle=
  \Gamma_0\,[\,0+0.92+0.46+\ldots]=\Gamma_0\,[1.71]
\end{equation}
{}From these expansions, a large contribution at
${\cal O}(\alpha_s^2\hat m_c^3\ln\hat m_c)$ can be seen. As discussed in the
introduction, one can expect a cancelation of this term by the
$\alpha_s^2\hat m_c^4\ln\hat m_c$ term, which has not been calculated here.

\section{Summary}

We have studied the operator product expansion for the process $b\to
c\ell\bar\nu$, to understand better the origin of the ``phase space''
logarithms which appear in the total decay rate.  After extracting the known
tree level term, we have extended the analysis to include radiative
corrections.\footnote{The one loop radiative corrections to the operator
product expansion for nonleptonic $D$ decays were calculated in
Ref.~\cite{BDS}, although no logarithms were resummed.}  In particular,
we have used a renormalization group analysis to resum the
leading logarithms of the form $\hat{m}_c^2\alpha_s^n\ln^n\hat{m}_c$,
$\,\hat{m}_c^3\alpha_s^{n+1}\ln^n\hat{m}_c$ and
$\,\hat{m}_c^4\alpha_s^n\ln^{n+1}\hat{m}_c$.  Unfortunately, these terms do not
dominate, in any limit of the theory, over certain others which have been
omitted.  Hence the results of this calculation cannot be used to extract any
reasonable estimate of the true size of the higher order corrections.

The point of this calculation lies rather in the insight which it affords us
into the origin of these logarithms, which even though not divergent, reflect
sensitivity to physics which is far in the infrared with respect to the scale
of the decaying $b$ quark.  We have exploited this separation of scales to
resum to all orders a certain subset of the phase space logarithms.  In so
doing, we have explored more generally their relation to other logarithms which
appear in the theory, such as the ``hybrid'' anomalous dimensions of the heavy
weak current~\cite{VS,PW}.  The hybrid anomalous dimensions are also not
numerically dominant, but by studying them one may investigate interesting
questions of principle in the Heavy Quark Effective Theory.  The analysis and
resummation which we have performed here should be viewed in much the same
spirit.

\acknowledgements

This work was supported by the United  States National Science Foundation
under Grant No.~PHY-9404057 and by the Natural Sciences and Engineering
Research Council of Canada.  A.F.~acknowledges additional support from
the United States National Science Foundation for National Young
Investigator Award No.~PHY-9457916, the United States Department of
Energy  for Outstanding Junior Investigator Award No.~DE-FG02-94ER40869
and the Alfred P.~Sloan Foundation.

\end{document}